\documentclass[pre,twocolumn,showpacs]{revtex4}
\begin{document}
\title{
Residence Time Statistics for Normal and  Fractional Diffusion
in a Force Field
}
\author{E. Barkai$^1$ }
\affiliation{$^1$Department of Physics,
Bar Ilan University, Ramat-Gan 52900 Israel}
\email{barkaie@mail.biu.ac.il}
\begin{abstract}
{
We investigate statistics of occupation times for an over-damped Brownian
particle in an external force field. 
A  backward Fokker-Planck equation introduced by
 Majumdar and Comtet describing the
 distribution of
occupation times is solved.
The solution gives a general relation between occupation time statistics
and probability currents which are found from solutions of the corresponding
problem
of first passage time. This general relationship between occupation times
and first passage times, is  valid for  normal Markovian
diffusion  and for non-Markovian sub-diffusion, 
the latter  modeled using the fractional 
Fokker-Planck equation. For binding potential fields we find
in the long time limit ergodic behavior for normal diffusion,
while for the fractional framework weak ergodicity breaking is 
found, in agreement with previous results of Bel and Barkai on the
continuous time random walk on a lattice. 
For non-binding potential rich physical
behaviors are obtained, and classification of occupation time
statistics is made possible according to whether or not the underlying
random walk is recurrent and the averaged first return time to the
origin is finite. 
Our work establishes a link between 
fractional calculus and ergodicity breaking. 
}
\end{abstract}
\maketitle

\section{Introduction}

 Consider the trajectory of a single Brownian particle.
The total time the particle spends in a given domain is called
the residence time or 
the occupation time.
The well known  example is P. L\'evy's arcsine law \cite{Satya,Redner}. 
Consider a Brownian motion $\dot{x}(t) = \eta(t)$ where
$\eta(t)$ is Gaussian white noise with zero mean. L\'evy investigated
the residence  time of the particle in
the domain $x>0$, which we call $T^{+}$,  when the motion is unbounded 
and  the total observation time is $t$.
Naive expectation is that 
$T^{+}/t = 1/2$ with small fluctuations when $t \to \infty$, namely
the particle occupies the domain $x>0$ for half of the
time  of observation. However,
instead the probability density function of  $T^{+}/t$ is given by
the well known arcsine law, 
$f(T^{+}/t)=[\pi\sqrt{ (T^{+}/t)(1 - T^{+}/t)}]^{-1}$
with $0 \le T^{+}/t\le 1$. This probability density
has a U 
shape, which means that for a typical realization of the Brownian
trajectory,  the particle spends most of the time in one half of space
(say $x>0$) and not in the other ($x<0$). 

 Many extensions of this well known result are found in the
literature. Darling and Kac \cite{Darling,Weiss}
found the limiting distribution
of the time spent in a domain in two dimensions, and this line of
investigation  was
extended to three dimension by Berezhkovskii et al \cite{Berez}. 
Lamperti's \cite{Lamp} limit theorem gives a very general
mathematical foundation for occupation time statistics (see more 
details in the manuscript). 
Recently in \cite{Benichou}  
Pearson's type of ballistic motion with random reorientation
was considered,  instead of the usual
assumption of an underlying continuum  process. 
The basic mathematical theory for the calculation of
occupation time statistics for Brownian motion
was developed by Kac, and is usually based on the
Feynmann-Kac formula (see \cite{Satya,Williams} and Ref. therein). 
Statistics of occupation times
is of-course not limited to Brownian motion and diffusion, 
and it is a topic
of wide investigation \cite{Satya}, for example in
the context of renewal processes \cite{Godreche}, 
theory and experiments of blinking quantum dots \cite{Gennady,Gennady1}, 
weak ergodicity breaking of dynamics generated using  
deterministic maps \cite{GolanMaps}, and work distribution functions
of a single spin \cite{Dhar1}. 

 The problem of occupation times of a Brownian
particle  in the presence of
external field was consider recently, by
Majumdar and Comtet \cite{Majumdar}. Using the Kac formalism \cite{Satya}
they found a backward
Fokker--Planck equation  whose solution yields statistics
of occupation times.  
In \cite{Majumdar} the problem  of occupation time
statistics of a particle performing a random walk
on a random walk i.e.  the Sinai model was investigated.
It was shown
that statistics of occupation times are drastically changed
when averages over random disorder are made. 

 In the first part  of this manuscript we consider the problem of
occupation times for normal Brownian motion
in an external fields. We solve
exactly the backward Fokker--Planck equation given in \cite{Majumdar}.
This solution gives a general  relation between occupation
time and first passage time statistics. Besides
the theoretical interest in such a relation, 
the solution is used to classify very general  behaviors of 
occupation times based on the corresponding 
properties of the first passage times. The later are investigated
in great detail in the literature \cite{Redner}, and we
can  use this knowledge to solve analytically the  
problem of  occupation times  at-least for some simple
cases.  For example we show
that in the limit of long measurement  times, and for binding force fields,
statistics of occupation times  
is determined by Boltzmann's statistics, namely the underlying
dynamics is ergodic, as expected. 

 Statistics of occupation time
is important from a fundamental point of view, since if
we are able to calculate statistics of occupation times
from some underlying dynamics, one can check the validity
of the ergodic hypothesis and its possible extensions. 
A trivial example is Gaussian Brownian
motion in a system of finite size $0<x<L$, 
in the absence of external force fields.
Then it is easy to show that
the  residence time in half of the system i.e. in the
domain $(0,L/2)$, is in statistical sense
half of the observation time,
when the observation time is long, 
as expected. For dynamics described by fractional kinetic
equations \cite{Metzler,Soko}, 
we show that such a simple ergodic picture does not
hold.  

 In the second part of the paper we consider the
problem of a particle undergoing an anomalous
diffusion process. We model this behavior using the
fractional time Fokker-Planck equation \cite{MBK,PRE2}. This
fractional framework  is based on fractional calculus e.g.
$d^{1/2} / d t^{1/2}$, which is briefly introduced in the manuscript.  
  We show
for example, that the  general relation between occupation
times and first passage times we find in the first
part of the paper is still valid, even
for the non-Markovian sub-diffusive case.
Similar to normal diffusion case a classification
of typical behaviors of occupation times
is found, and analytical solutions provided.
For dynamics in binding force fields we find weak ergodicity breaking. 
In the conclusions we compare our results on occupation times
found here using the fractional framework, and recent results
of Bel and Barkai \cite{Golan,Golan1} on statistics of residence times
for continuous time random walks.  


For applications, residence times are of interest
in the context of chemical reactions \cite{Blumen,Agmon,BarHaim} 
and rather generally
for statistical analysis
of experimental data. 
 Residence times are very important in the context of
single molecule dynamics \cite{BJS,Bustamante}. 
 It is now possible to follow dynamics of single molecules
embedded in condensed phase environments, using optical
techniques. 
For example dynamics of single molecules in cells or in solution
are used to follow chemical reactions in real time, without
the problem of ensemble averaging found in usual measurements.
A typical experiment uses a laser to investigate the
dynamics of a particle. In many cases and under certain
conditions \cite{BJS} if a particle or a reaction
coordinate 
is in a finite domain, the system may emit photons, while when
the particle is out of the domain the system does not emit. 
Very briefly,  the domain width can
be imagined as the width of the laser beam in single molecule
fluorescence experiments when the particle comes in and out
of resonance with the exciting laser field, due to its diffusion
in space e.g. \cite{Zumofen}, 
 or it could be  the F$\ddot{o}$ster 
radius in fluorescence resonance energy transfer measurement e.g. \cite{Haran}.
Thus the total time the photons are emitted
is approximately the residence time, which is proportional to the number of
emitted photons,  which is generally a random variable. 
For other sources of fluctuations in single molecule
experiments see \cite{BJS}.

\section{Normal Diffusion}

 Consider a one dimensional
 over-damped Brownian motion in an external force
field $F(x)$. The Smoluchowski Fokker--Planck equation for
the concentration of non interacting particles is
\begin{equation}
{\partial c(x,t) \over \partial t} = D \left[
{\partial^2 \over \partial x^2} - {\partial \over \partial x} { F(x) \over k_b T} \right] c(x,t),
\end{equation}
where $T$ is the temperature and 
$D$ is the diffusion coefficient. As well known the equilibrium
of the ensemble of particles is the Boltzmann equilibrium, provided that
the force field is binding.  

Consider a single particle, which 
at time $t=0$ is on $x_0$, 
the observation time of the stochastic dynamics is $t$.  
The random variable we investigate
here is  
$T^{+}$, the total time the particle occupies the region
$x>0$. In principle  during the observation time the particle may cross
the point $x=0$ many times, and then the occupation time
$T^{+}$ is composed of many
sojourn times in $x>0$.

 Let $P_{x_0,t}(T^{+})$ be the probability density function (PDF) of
$T^{+}$. The double Laplace transform 
\begin{equation}
P_{x_0,s}\left(u \right) = \int_0 ^{\infty} \int_0 ^{\infty} e^{ - s t} e^{- u T^{+} } P_{x_0,t}(T^{+}) {\rm d} t {\rm d} T^{+}, 
\label{eq01}
\end{equation}
is defined so that $s$ and $t$ and $u$ and $T^{+}$ are Laplace pairs.
Majumdar and Comtet \cite{Majumdar} found the equation of motion for
$P_{x_0,t}(T^{+})$
in double Laplace space
\begin{equation}
 D \left[ {\partial^2 \over  \partial x_0^2} +
{ F\left( x_0\right) \over k_b T} {\partial \over \partial x_0} \right] P_{x_0,s}\left(u\right) - 
\left[s + \Theta\left(x_0\right) u\right] P_{x_0,s}\left(u \right)
= -1. 
\label{eq02}
\end{equation}
Where 
$\Theta(x_0)$ is the step function:
$\Theta(x_0)=1$ if $x_0>0$ otherwise it is zero.
This type of equation is called a
backward Fokker--Planck equation,  
the operator on the left hand side depends on the initial
condition $x_0$. Eq.
(\ref{eq02})
is solved for the matching  boundary conditions
$$ P_{x_0,s}(u)|_{x_0 = 0^{+}}
=P_{x_0,s}(u)|_{x_0 = 0^{-}},$$
\begin{equation}
{\partial P_{x_0,s}(u) \over \partial x_0} |_{x_0 = 0^{+}}
={\partial P_{x_0,s}(u) \over \partial x_0} |_{x_0 = 0^{-}}. 
\label{eq04}
\end{equation}
We will re-derive Eq.  
(\ref{eq02}) later as a special limiting case of a more general
non-Markovian dynamics. 

To prepare for the solution
of Eq. (\ref{eq02})
we define the following survival probabilities.
The probability that a  particle starting at $x_0$ with $x_0<0$,  
to remain  in the domain $x<0$ without leaving it even once, 
during the time
$t$ is the survival probability $W^{-}_{x_0}(t)$.
Let 
$W^{-} _{x_0}(s)$ be the Laplace transform of $W^{-}_{x_0}(t)$
and 
similarly the Laplace transform of the survival probability
in the domain $x>0$ is
$W^{+} _{x_0}(s)$ for $x_0>0$.
The key to the solution of the problem of occupation
times in half space,  is to recall
Tachiya's  equation for the survival probability of
a particle in half space \cite{Tachiya,SatyaAlain} 
\begin{equation}
D \left[ {\partial^2 \over\partial x_0^2} + { F\left( x_0\right) \over k_B T} {\partial \over \partial x_0} \right] W^{-} _{x_0}\left(s\right) - s  W^{-} _{x_0}\left(s\right)
= -1  \ \ \ x_0 <0,
\label{eq07}
\end{equation}
and a similar equation holds for $x_0>0$.
The boundary conditions for Eq. (\ref{eq07}) 
are the standard conditions used for the calculation of
survival probabilities. Namely, $W^{-}_{x_0}(s)|_{x_0=0}=0$, 
means that the particle 
reaches the boundary on $x=0$ instantaneously if the particle
starts very close to the absorbing boundary and if
$x_0 \to -\infty$ survival is unity. 

The solution of Eq. (\ref{eq02}) for the occupation time is
$$ P_{x_0,s}\left( u \right)= W^{-} _{x_0} (s) + \left[ 1 - s W^{-} _{x_0}(s) \right] G_s(u)$$
if  $x_0<0$,
\begin{equation}
P_{x_0,s}\left( u \right) = W^{+} _{x_0} (s+u) + \left[ 1 - (s+u) W^{+} _{x_0}(s+u) \right] G_s (u)  
\label{eq06}
\end{equation}
if  $x_0>0$.
From Eq. (\ref{eq06})
the physical meaning of
$G_s(u)$ becomes clear, it  is the double Laplace transform
of $G_t(T^{+})$ the PDF of the random variable 
$T^{+}$ for a particle starting on $x_0=0$. 
The PDF $G_t(T^{+})$ contains the information
on the problem of occupation times, while the survival probability
was investigated previously by many authors, hence in what follows
we investigate $G_t(T^{+})$. 
Using Eq. (\ref{eq07}) 
the reader can verify that  
Eq. (\ref{eq06}) is indeed the general solution
of the problem of occupation times Eq. 
(\ref{eq02}).
Using the boundary condition 
$W^{-}_{x_0 =0}(s)=W^{+}_{x_0 =0}(s)=0$  
and the solution Eq.
(\ref{eq06})
it is easy to see that
the boundary condition
$P_{x_0,s}(u)|_{x_0 = 0^{+}}
=P_{x_0,s}(u)|_{x_0 = 0^{-}} = G_s(u)$ in Eq. (\ref{eq04}) is satisfied. 

The second matching boundary condition in Eq. (\ref{eq04}),
on the derivatives of $P_{x_0,s}(u)$
yields  $G_s(u)$ 
using
Eq. (\ref{eq06})  
\begin{equation}
G_s(u) = {  J^{+}(s + u ) -J^{-} (s) \over 
(s + u ) J^{+} (s + u) - s J^{-} (s) } 
\label{eq08}
\end{equation}
where the currents are
\begin{equation}
J^{+} (s + u) = { \partial W^{+} _{x_0} (s+u) \over  \partial x_0} |_{x_0=0^{+} }, 
\ \
J^{-} (s) = { \partial W^{-} _{x_0} (s) \over \partial x_0} |_{x_0=0^{-}}. 
\label{eq09}
\end{equation}
Eqs. (\ref{eq06},
\ref{eq08}) are the main results so far since they yield
a general relation between statistics of occupation
times and survival probability currents. 
From Eq. (\ref{eq09}) we see that
the solution of the problem of occupation times is found
in terms of two solutions of the  corresponding first passage time
problems, the first for a particle starting on $x_0>0$ and absorbed 
on $x=0$ (i.e. $J^{+}$) and the 
second for a particle starting on $x_0<0$ and
absorbed on $x=0$ (i.e. $J^{-}$). 
Thus the problem of residence times is solved in three steps:\\
(i) Find solutions of two first passage time problems
for, $x_0>0$ and $x_0<0$ in Laplace space. \\
(ii) Use Eq. (\ref{eq08}) to find the solution of the problem
of residence times in double Laplace space. \\
(iii) And then use a two dimensional inverse Laplace transform
to get $G_t (T^{+})$ from $G_s(u)$.\\
Since there exists a vast literature
on the  solutions of the
problem of first passage time \cite{Redner}, 
the 
relationship Eq. 
(\ref{eq08})
is very useful for the calculation of statistics of occupation
times. 
We note that some connections between first passage times
and occupation times, which are different and in our
opinion less general than Eq. (\ref{eq09}), appeared previously in the
literature \cite{Berez,Williams,Golan1}. 
Finally, while we considered the 
occupation time in half space, occupation times in
a finite domain are also obtained in a similar way,
and it is straight forward
to extended our results to higher dimensions.

Majumdar and Comtet \cite{Majumdar} classify statistics of occupation
times according to behavior of the potential field, in particular
they consider motion in stable, unstable and flat potential fields.
Here
the relation between survival currents and statistics of
occupations times, Eq. (\ref{eq08}) can be used to characterized
certain very general and new behaviors of occupation times.

Survival probabilities in a finite and infinite  domain exhibit
three well known typical physical behaviors \cite{Redner}, 
we consider the right
random walk (i.e. $x_0>0$) and similar classification holds for the
left random walk. Later we will classify behaviors
of residence times based on these three behaviors of first passage times. \\
{\bf Case 1} The random walk is recurrent, 
and the average first passage time from $x_0$ to $0$
is finite. Such cases correspond to diffusion in a system
of finite size, when the particle cannot escape to infinity,
e.g. the driving force field is binding.\\
{\bf Case 2} The random walk is transient, i.e. the survival probability
in $x>0$ is finite in the limit of long times. Such cases
happen when the external force drives the particle
far from the origin, and the system is infinite.
In that case in the limit of small $s$
\begin{equation}
W_{x_0} ^{+} \sim { {\cal \epsilon}^{+} _{x_0}  \over s}
\label{eq11a}
\end{equation} 
where ${\cal \epsilon}^{+} _{x_0}$ is the survival probability
of the particle starting on $x_0$, without reaching
$x=0$, when $t \to \infty$. Similar notation is used for the left
random walk, with ${\cal \epsilon}^{-} _{x_0}$. 
 \\
{\bf Case 3} Random walks are recurrent, though the average first passage time
is infinite. A particularly common situation is the case
when the survival probability decays like $t^{-1/2}$ for long times.
This happens if the non-diverging external field $F(x)=0$ for $x>x_c$ and the
system is infinite, namely when diffusion controls the long
time dynamics. 
For such a  case \cite{Redner}
\begin{equation}
W_{x_0} ^{+} \sim { A^+ _{x_0} \over s^{1/2}}, \ \ s\rightarrow 0,
\label{eq11b}
\end{equation} 
where  $A^+ _{x_0}>0$ depends of-course
on the details of the force field.\\
We now  consider certain general properties
of the statistics of occupation times for the three
cases.  

{\bf Case 1} The long time behavior of $G_t(T^{+})$ is now considered. 
When the left and right 
random walks, starting
at  $x_0<0$ or $x_0>0$, respectively, are recurrent and the average first
passage time is finite. For this case 
the small
$s$ limit yields
\begin{equation}
W^{\pm} _{x_0} (s=0) = \langle t^{\pm} _{x_0} \rangle,
\label{eq10}
\end{equation} 
where $\langle t^{\pm} _{x_0} \rangle$ is the average time for the
particle starting on $x_0<0$ (or $x_0>0$) 
to reach the origin for the first time.
The small $s$ and $u$ limit, their ratio arbitrary, of Eq.
(\ref{eq08}) gives the long $t$ and $T^{+}$ behavior of
$G_t(T^{+})$, we find
\begin{equation}
G_s (u) \sim { 1 \over s + u {
{\partial \langle t^{+}_{x_0} \rangle \over \partial x_0}|_{x_0=0} \over 
{\partial \langle t^{+}_{x_0}\rangle \over \partial x_0}  |_{x_0=0}-
{\partial \langle t^{-}_{x_0} \rangle \over \partial x_0} |_{x_0=0}} }.
\label{eq11}
\end{equation}
The differential equation for $\langle t^{+}_{x_0} \rangle$ is well known
and is obtained from the small $s$ expansion of 
Eq.
(\ref{eq07})
\begin{equation}
D\left[ {\partial^2 \over \partial x_0^2 } \langle t^{+} _{x_0} \rangle +
{F(x_0) \over k_B T } {\partial \over \partial x_0} \langle t^{+}_{x_0} \rangle \right] = - 1.
\label{eq12}
\end{equation}
Solving this equation, using a similar equation
for $\langle t^{-} _{x_0} \rangle$,  and inverting 
Eq. (\ref{eq11})
to the time domain
we find the expected ergodic behavior
\begin{equation}
G_t ( T^{+} ) \sim \delta\left( T^{+} - P_B ^{+} t \right),
\label{eq13}
\end{equation}
where $P_B ^{+}$ is Boltzmann's probability
of occupying $x>0$
\begin{equation}
P_B ^{+ } = {\int_0 ^\infty e^{ - { U(x) \over k_B T  } } {\rm d} x 
\over Z},
\label{eq14}
\end{equation}
$Z = \int_{-\infty}^\infty \exp - { U(x)\over k_B T  }  {\rm d} x$
is the normalizing partition function and $U(x)$ is the binding potential,
with $F(x) = - {\rm d} U(x) / {\rm d} x$. 

{\bf Case 2} We consider the case where both the left and the right
random walks are non recurrent. The survival probabilities in the
two domains are ${\cal \epsilon}^+ _{x_0}$ and 
${\cal \epsilon}^- _{x_0}$, in the long time limit.  Then using  
Eqs. (\ref{eq08},\ref{eq11a}) we find for $t \to \infty$
\begin{equation}
G_t(T^{+}) \sim \alpha^{-} \delta(T^{+}) + \alpha^{+} \delta(T^{+} - t)
\label{eq14a}
\end{equation}
where
\begin{equation}
 \alpha^{+} = {  { \partial {\cal \epsilon}^{+} _{x_0} \over \partial x_0} |_{x_0 = 0} \over 
{ \partial {\cal \epsilon}^{+} _{x_0} \over \partial x_0} |_{x_0 = 0}  -
{ \partial {\cal \epsilon}^{-} _{x_0} \over \partial x_0} |_{x_0 = 0}  }
\label{eq14b}
\end{equation}
and $\alpha^- = 1 - \alpha^+$.
%
%
Since the particle always manages to escape either to
the left or to the right, eventually the particle will
either reside in the left domain or the right domain forever, hence the
delta functions in Eq. (\ref{eq14a}). The weights of these delta
functions are given by the derivatives of the survival probabilities
only. 

{\bf Case 3} We now consider a case where
 both the left and the right random walks
are recurrent, though the average first return time
from $x_0$ to $x=0$ is infinite, in such a way that Eq.
(\ref{eq11b})
is valid. Then in the small
$s$ and $u$ limit
\begin{equation}
G_s(u) \sim { s^{ - 1/2} + {\cal R} (s + u)^{-1/2} \over 
s^{1/2} + {\cal R} (s + u)^{1/2}}
\label{eqGSU}
\end{equation}
where the asymmetry parameter is
\begin{equation}
{\cal R} = - { {\partial A^{+}_{x_0} \over \partial x_0}|_{x_0 = 0} \over
 {\partial A^{-}_{x_0} \over \partial x_0}|_{x_0 = 0} }.
\label{eqAsym}
\end{equation}
Transforming to the time domain we find
the asymmetric arcsine PDF \cite{Lamp}
\begin{equation}
G_t (T^{+} ) \sim {1 \over t} f\left({T^{+} \over t} \right)
\label{EqLamh}
\end{equation}
where
\begin{equation}
f(x) = { 1 \over \pi} { {\cal R} \over x^{1/2} \left( 1 - x\right)^{1/2}\left[ {\cal R}^2 \left( 1 - x \right) + x \right]}. 
\label{eqGAR}
\end{equation}
When ${\cal R} = 1$ we find the  arcsine law. 
Note that
the PDF Eq. (\ref{EqLamh}) diverges on $T^{+}/t = 1$ and
$T^{+}/t = 0$, hence events where the particle always occupies
(or hardly never occupies) the domain $x>0$ 
have a significant contribution.

Another general result obtained from the small $u$ expansion
of  Eq. (\ref{eq08}) is for the
average occupation time
\begin{equation}
\langle T^{+} \rangle = {\cal L}^{-1} _{s \to t} \left\{ 
{  J^{+} (s) \over s^2 \left[J^{+}(s) - J^{-} (s) \right] } \right\}
\end{equation}
where ${\cal L}^{-1} _{s \to t}$ is the inverse Laplace transform.
If the potential field is binding and the random walk is
recurrent, then $\langle T^{+} \rangle \sim P_B ^{+} t$. 
Similar relations between higher order moments
of the occupation times and the survival probabilities are
obtained in a similar way. 
In Sec.  \ref{SecExa} we consider several particular examples,
which explain in greater detail the meaning of the
general results obtained in this section. 
First we generalize our results for  fractional
dynamics. 

\section{Anomalous Diffusion}

 Anomalous diffusion and relaxation is modeled based
on the fractional time Fokker--Planck equation (FFPE) \cite{MBK,PRE2},
the concentration of non-interacting particles obeys
\begin{equation}
{\partial^{\alpha}  c(x,t) \over \partial t^\alpha} = D_{\alpha} \left[
{\partial^2 \over \partial x^2} - {\partial \over \partial x} { F(x) \over k_b T} \right] c(x,t),
\label{eqFFPE}
\end{equation}
where $D_{\alpha}$ is a generalized diffusion coefficient
and $0<\alpha<1$.
A brief mathematical  introduction to the FFPE is given in Appendix A. 
We recall physical properties of the FFPE. (i) when $F(x)=0$ 
and for free boundary conditions we have the fractional
diffusion equation \cite{Bala,Schneider,Saichev,BarCP,Gor} 
with anomalous diffusion
$\langle x^2 \rangle \propto t^{\alpha}$. (ii) In
the presence of a binding time independent force field
the equilibrium is the Boltzmann distribution \cite{MBK,PRE2}. (iii)
Generalized Einstein relations are satisfied
in consistency with linear response theory \cite{MBK,PRE2}. (iv) Relaxation
of modes follows the Mittag Leffler decay, related for example
to Cole-Cole relaxation \cite{MBK,PRE2}.  (v) In the limit $\alpha \to 1$ 
we recover the standard Smoluchowski Fokker-Planck equation. 
The FFPE is derived from the continuous time random walk \cite{PRE2}.
Its mathematical foundation is P. L\'evy's generalized central limit theorem
applied to the number of steps in the underlying
random walk \cite{PRE1,Meer}. A very general solution of the
FFPE in terms of the solution
of the standard $\alpha=1$  Fokker-Planck equation was given
in \cite{PRE1}
(i.e., subordination, and the inverse L\'evy transform).
Recently there is some controversy on how to apply
boundary conditions 
\cite{Sung,Seki,Chechkin} for the anomalous case. 
Applications of fractional diffusion modeling include:
Scher-Montroll time of flight transport of charge carriers
in disordered medium \cite{PRE1},
dynamics of ion channels \cite{Hanngi}, relaxation processes in
proteins \cite{Kneller}, and dielectric relaxation
\cite{Coffey}, and deterministic diffusion \cite{Saichev}. 
For a review and a popular article on fractional kinetics
see \cite{Metzler,Soko}.

Similar to the normal diffusion case, we define $P_{x_0,t}(T^{+})$ as
the PDF of $T^{+}$ and $P_{x_0,s} (u)$, its double Laplace transform.
As we show in next subsection  
the differential Eq.  for $P_{x_0,s} (u)$ for the dynamics
described by the FFPE Eq. (\ref{eqFFPE})  is
$$ D_{\alpha} \left[ {\partial^2 \over  \partial x_0^2} +
{ F\left( x_0\right) \over k_b T} {\partial \over \partial x_0} \right] P_{x_0,s}\left(u\right) - $$
\begin{equation}
\left[s + \Theta\left(x_0\right) u\right]^{\alpha} P_{x_0,s}\left(u \right)
= - \left[s + \Theta\left(x_0\right) u\right]^{\alpha-1}.
\label{eqfrac02}
\end{equation}
The boundary conditions for Eq.
(\ref{eqfrac02})
 are identical to the normal diffusion
case $\alpha=1$ given  in Eq. 
(\ref{eq04}). 
Eq. (\ref{eqfrac02}) is important since as pointed out in
\cite{Golan},  fractional dynamics is
weakly non-ergodic \cite{Bouchaud}, namely occupation time statistics
is not described by Boltzmann equilibrium even in the limit of 
long time and for binding potential fields. Thus the FFPE 
(\ref{eqFFPE})
cannot be used to describe time
averages of physical observable due to ergodicity breaking, 
and the interpretation of results derived from the FFPE
must be treated with care.
Eq.
(\ref{eqfrac02}) is a remedy for this problem since as we will show
it yields a fractional framework for the calculation of
non-trivial distribution of occupation times [i.e.
generalization of Boltzmann's statistics Eq. 
(\ref{eq13})].
Eq. (\ref{eqfrac02}) is a fractional backward Fokker--Planck equation
in double Laplace space, formally one may invert it to the time
domain using material fractional derivatives \cite{sokoMet}, however in practice
we solve this equation in double Laplace space and only then
make the inverse double Laplace transform. 

Interestingly the solution of the fractional Eq. (\ref{eqfrac02}) 
is identical to
that found for normal diffusion case, namely our main results Eqs.
(\ref{eq06},\ref{eq08},\ref{eq09})
are valid also in the non-Markovian domain $0<\alpha<1$. 
Now $W_{x_0}^{\pm}(s)$ needed for the calculation
of $J^{\pm}(s)$, is the  Laplace transform of the survival
probability for the  fractional
particle.
Thus Eqs. (\ref{eq06},\ref{eq08},\ref{eq09}) 
have some general validity
beyond normal Markovian diffusion.  

To prove  that 
Eqs. (\ref{eq06},\ref{eq09}) are still valid  
we must first find the differential  equation
for  $W_{x_0}^{+} (s)$:
the survival probability of a fractional
particle starting on $x_0>0$ in the domain $x>0$.  
We can prove that 
\begin{equation}
D_{\alpha} \left[ {\partial^2 \over\partial x_0^2} + { F\left( x_0\right) \over k_B T} {\partial \over \partial x_0} \right] W^{+} _{x_0}\left(s\right) - s^{\alpha}  W^{+} _{x_0}\left(s\right)
= -s^{\alpha -1},
\label{eqfrac07}
\end{equation}
and a similar equation holds for $x_0<0$.
Eq. (\ref{eqfrac07}) is the fractional generalization of Tachiya's
Eq. (\ref{eq07}). The derivation of Eq. (\ref{eqfrac07}) 
is based on results obtained in  \cite{PRE1} and
is simple once the sub-ordination trick is used 
(see some details in Appendix A). 
Now using Eq. (\ref{eqfrac07}) it is easy to 
verify that Eqs. 
(\ref{eq06},\ref{eq09}) are solutions of the fractional
Eq. (\ref{eqfrac02}).

\subsection{Derivation of Fractional Equation for Occupation Times}

 In this subsection we derive our main result Eq. 
(\ref{eqfrac02})
using
the assumption that the underlying dynamics is described by
the fractional Fokker-Planck equation 
(\ref{eqFFPE}). The latter describes
long time behavior of the continuous time random
walk (CTRW), which is the underlying  random walk process we have in mind.
In the CTRW a particle
performs a one dimensional random walk on a lattice, with jumps to
nearest neighbors only and with random waiting times
between jumps. In CTRW the waiting times between jumps
are independent identically distributed random variables, namely
the CTRW  process is renewed after each jump.
The PDF of waiting times is $\psi(t)$. Two classes
of CTRWs are usually considered, the case when the average waiting
time is finite, and the case when $\psi(t) \propto t^{ - (1 + \alpha)} $
when $t \to \infty$ 
and $0<\alpha<1$. The latter case leads to a non-stationary behavior, 
aging,  anomalous diffusion and weak ergodicity breaking \cite{Bouchaud}. 
The lattice spacing is $\epsilon$, eventually we will
consider the continuum limit when $\epsilon$ is small.
On each lattice point we assign a probability for jumping
left and a probability of jumping right. 
This dynamics in the continuum limit leads to behavior described
by the FFPE, when detailed balance conditions are
applied on the probabilities for jumping left or right \cite{PRE2} (i.e.
probabilities to jump left or right are related to external force field
and temperature). 
In what follows
we start with some general arguments, assuming only a
renewal property of the random walk,  without limiting our selves to
a specific model. 

The random position of the particle is $x(t)$. 
The total time the particle spends on $x \ge 0$ is $T^{+}$,
i.e. the occupation time of half space. The particle starts
on $x_0$ and assume that  $x_0\ge 0$,
later we generalize our results to the case $x_0<0$.  
We define the PDF of first passage times, from $x_0$ to
$x=-\epsilon$, as $\psi_{x_0} ^{+}(t)$. 
 The PDF of first passage
times from $x=0$ to $x=-\epsilon$ ($x=-\epsilon$ to $x=0$) is denoted with
$\psi^{+}(t)$ and $[\psi^{-}(t)]$
respectively.

 We assume that the first passage
times PDFs $\psi^{+}(t)$ and $\psi^{-}(t)$ do not depend on 
$x_0$ and that sojourn times in domain $x>0$ and 
$x<0$ are statistically independent. 
Such assumption holds for Markovian dynamics
but is not obvious otherwise. For CTRW dynamics the assumption
is correct, since as mentioned the CTRW processes is a renewal process.
 The process is mapped on a two
state process
\begin{equation}
\theta_x(t) = \left\{
\begin{array}{c c} 
1 & \ x(t) \ge 0 \\
0 & \ x(t) < 0
\end{array}
\right.
\end{equation}
and hence 
$ T^{+} = \int_0 ^t \theta_x(t) {\rm d} t$. Since either the particle
is in the domain $x<0$ or not the dynamics
is described by a set of sojourn times
$$ \tau_{x_0},\tau_1 ^{-},\tau_2 ^{+},\tau_3 ^{-}, \tau_4 ^{+} \cdots.$$
Here the PDF of $\tau_{x_0}$ is $\psi_{x_0}(t)$,
the PDF of $\tau_1 ^{-}$ is $\psi^{-}(t)$, the PDF  of $\tau_2 ^{+}$
is $\psi^{+}(t)$, etc. All the sojourn times are
assumed mutually independent, which means the process is renewed once
the particle jumps from $x=0$ to $x=-\epsilon$ or vice
versa. 

 Let $f_{x_0,t}(T^{+})$ be the PDF of $T^{+}$ when the total
observation time is $t$. Let $f_{x_0,s}(u)$ be the double Laplace
transform of $f_{x_0,t}(T^{+})$. A calculation, using methods
of renewal theory, 
similar to the work of Godreche and Luck \cite{Godreche}
yields
\begin{widetext}
\begin{equation}
 f_{x_0,s} (u) =  
{ 1 - \psi_{x_0} ^{+}(s + u) \over s + u} + 
\psi_{x_0}^{+} (s + u)
\left[ \psi^{-} (s) { 1 - \psi^{+} (s + u ) \over s + u}  + {1 - \psi^{-}(s) \over s} \right] { 1 \over 1 - \psi^{+}(s+u) \psi^{-} (s) }.
\label{eqLong}
\end{equation}
Where 
$\psi_{x_0} ^{+}(s + u)= \int_0 ^t \exp[ - (s + u) t] \psi_{x_0} ^{+}(t) {\rm d} t$ is the Laplace transform. 
If $\psi_{x_0} ^{+}(t)=\psi^{+}(t)=\psi^{-} (t)$ we recover a result
in \cite{Godreche}.
If the particle starts on $x_0<0$ then one can show
\begin{equation}
 f_{x_0,s} (u) =  
{ 1 - \psi_{x_0} ^{-}(s) \over s } + 
\psi_{x_0}^{-} (s) 
\left[ \psi^{+} (s + u) { 1 - \psi^{-} (s  ) \over s }  + {1 - \psi^{+}(s +u) \over s + u } \right] { 1 \over 1  - \psi^{+}(s+u) \psi^{-} (s) }.
\end{equation}

  We now consider the case when the underlying dynamics is
described by the FFPE.
  By definition the first passage time PDFs are related to survival
probabilities according to
\begin{equation}
W_{x_0} ^+ (t) = 1-  \int_0 ^t \psi_{x_0} ^+ (t) {\rm d} t,
\end{equation}
or using the convolution theorem in Laplace space
\begin{equation}
W_{x_0} ^+ (s) = {1 - \psi_{x_0} ^+ (s) \over s}.
\label{eqCON}
\end{equation}
Hence we may rewrite Eq. 
(\ref{eqLong})
\begin{equation}
 f_{x_0,s} (u) =  
W_{x_0}^{+} (s + u)
+ \left[ 1 - (s+u) W_{x_0} ^+ (s+u)\right]
\left[ \psi^{-} (s) { 1 - \psi^{+} (s + u ) \over s + u}  + {1 - \psi^{-}(s) \over s} \right] { 1 \over 1 - \psi^{+}(s+u) \psi^{-} (s) }.
\label{eqLong2}
\end{equation}
\end{widetext}
Notice that $f_{x_0,s}(u)$ depends on $x_0$ only through the survival
probability $W_{x_0} ^+ (s+u)$. If we  apply the backward Fokker-Planck operator
$$ D_{\alpha} \left[ {\partial^2 \over\partial x_0^2} + { F\left( x_0\right) \over k_B T} {\partial \over \partial x_0} \right] $$ 
on this equation and use Eq. (\ref{eqfrac07}), 
namely we  assume
that the underlying dynamics is described by the FFPE in the continuum limit, 
we obtain at once our main result Eq. 
(\ref{eqfrac02}). 
Similar method is used for the case $x_0<0$
to complete the proof. 

 We now derive our main Eq. 
(\ref{eq08})
using the continuum approximation.
We consider the case when the particle start on $x_0=0$ hence 
we have $\psi_{x_0} ^+ = \psi^+$ and define
\begin{equation}
\overline{G}_s(u) = f_{x_0 = 0, s} (u).
\end{equation}
Generally $\overline{G}_s(u)$  is not identical 
to $G_s(u)$ and  our aim now is to find the conditions
when these two functions are identical. 
 Using
(\ref{eqLong}) we have
$$ \overline{G}_s (u) = 
\left[ {1 - \psi^+(s+u) \over (s + u) } + \right.$$
\begin{equation}
\left.  \psi^+ (s + u) { 1 - \psi^- (s) \over s} \right] { 1 \over 1 - \psi^{+}(s+u) \psi^{-}(s) } .
\label{eqfsu7}
\end{equation}
In the continuum limit we have the following $\epsilon$ expansion
\begin{equation}
\psi^-_{\epsilon} (s) \simeq  \psi^-_{\epsilon =0} (s) - {\partial \psi^-_{\epsilon} \over \partial x}|_{\epsilon=0} \epsilon + \cdots 
\label{eqepsExp}
\end{equation}
and a similar expansion holds for $\psi^+ _{\epsilon} (s)$.
Where the subscript $_\epsilon$ 
in Eq. (\ref{eqepsExp})
is added to emphasize that the PDF  
of the first passage time  from lattice  point $-\epsilon$ to the origin $0$.
Note that $\psi^-_{\epsilon =0} (s)=1$, since the particle on the origin
is immediately absorbed. Inserting the expansion  
(\ref{eqepsExp}) in Eq. 
(\ref{eqfsu7})
and using a similar expansion for $\psi^+_{\epsilon } (s+u)$ we find
that when $\epsilon \to 0$
\begin{equation}
\overline{G}_s (u) \sim { {1 \over (s+u) } { \partial \psi^{+}(s+u) \over \partial x_0}|_{x_0=0} -  
 {1 \over s} { \partial \psi^{-}(s) \over \partial x_0}|_{x_0=0}  \over
 {\partial \psi^{+}(s+u) \over \partial x_0}|_{x_0=0} -  
  {\partial \psi^{-}(s) \over \partial x_0}|_{x_0=0} }
\label{eqbbb}
\end{equation}
Using Eqs. (\ref{eqCON},\ref{eqbbb}) we obtain our main result
Eq. 
(\ref{eq08}) and $\overline{G}_s(u)$ is identical to $G_s(u)$
in the continuum limit of $\epsilon  \to 0$.

\section{Examples}
\label{SecExa}

\subsection{Weak Ergodicity Breaking} 

 We now consider anomalous dynamics in a binding potential field $U(x)$, e.g. 
$U(x)= k x^2 / 2$ with $k>0$, and $F(x) = - {\rm d} U(x)/ {\rm d} x$. 
As we showed already the long time behavior of $G_t(T^{+})$, for normal
diffusion, yields an ergodic behavior Eq. 
(\ref{eq13}). Residence time statistics  for sub-diffusion in a binding
potential field was considered previously in \cite{Golan}, using
the continuous time random walk model on a lattice.
Here we consider a fractional Fokker-Planck equation approach
showing that concepts of weak ergodicity breaking in \cite{Golan}
are valid
also  within the fractional framework. 

 In \cite{PRE1} it was shown 
that the random walks in
a binding field are recurrent  for $\alpha<1$, 
just like the normal
case $\alpha=1$. Using the subordination trick (see Appendix), 
or analyzing Eq.
(\ref{eqfrac07})
we find that for $s \to 0$
\begin{equation}
W_{x_0} ^{\pm} (s) \sim { \left( \tau_{x_0} ^{\pm} \right)^\alpha \over s^{1 - \alpha} }.
\label{eqWEB01}
\end{equation} 
When $\alpha =1$ we have 
$\left( \tau_{x_0} ^{\pm} \right)^\alpha= \langle t_{x_0} ^{\pm} \rangle$ and 
the behavior in Eq.
(\ref{eq10}). For $0<\alpha<1$ we have in the time domain 
$W_{x_0} ^{\pm} (t)\propto t^{ - \alpha}$, reflecting the long tailed
trapping times of the underlying CTRW. The $\left( \tau_{x_0} ^{\pm} \right)^\alpha$ are amplitudes which satisfy 
\begin{equation}
D_{\alpha} \left[ { \partial^2 \over \partial x_0 \ ^2 } + { F(x_0) \over k_B T} {\partial \over \partial x_0} \right] \left( \tau_{x_0} ^{\pm} \right)^\alpha= -1.
\label{eqWEB02}
\end{equation}
This equation is obtained from the small $s$ expansion of Eq. 
(\ref{eqfrac07}).
Using Eq. (\ref{eqWEB01})
and Eq. (\ref{eq08}) we find in the limit of small
$s$ and $u$
\begin{equation}
G_s (u) \sim { {\cal R} (s + u )^{\alpha -1} + s^{\alpha -1} \over {\cal R} ( s + u)^\alpha + s^\alpha} 
\label{eqWEB03}
\end{equation}
where the asymmetry parameter is
\begin{equation}
{\cal R} = - { { \partial \left( \tau^{+} \right)^\alpha \over \partial x_0} |_{x_0 = 0^+} \over 
{ \partial \left( \tau^{-} \right)^\alpha \over \partial x_0 } |_{x_0 = 0^{-} } }.
\label{eqWEB04}
\end{equation}
Inverting to the time domain, we  see
that the PDF of $T^{+}$ in the long time $t$ limit is described 
by Lamperti's limit theorem \cite{Lamp}
\begin{equation}
G_t \left( T^{+} \right) \sim { 1 \over t} \delta_{\alpha} \left( {\cal R}, {T^{+} \over t} \right).
\label{eqWEB05}
\end{equation}
where the scaling function is
$$ \delta_{\alpha} \left( {\cal R} , p \right)  \equiv $$
\begin{equation}
{ \sin \pi \alpha \over \pi } 
{ {\cal R} p^{\alpha - 1} \left( 1 - p \right)^{\alpha -1} \over
{\cal R}^2 \left( 1 - p\right)^{2 \alpha} + p^{2 \alpha} +  2 {\cal R} \left( 1 - p\right)^{\alpha} p^{\alpha} \cos \pi \alpha }.
\label{eqWEB06}
\end{equation}
This function is normalized according to $\int_0 ^1 
\delta_{\alpha} \left( {\cal R} , p \right) {\rm d} p = 1 $.
When $\alpha =1$ we find the ergodic behavior in 
Eq. (\ref{eq13}), while clearly if $\alpha<1$ we find a non-ergodic behavior. 
The parameter ${\cal R}$ is called the asymmetry parameter.
It can be calculated solving Eq. 
(\ref{eqWEB02}), we find
\begin{equation}
{\partial \left( \tau_{x_0} ^{+}  \right)^{\alpha} \over \partial x_{0}} |_{x_0 = 0^{+}}=
{ 1 \over D_{\alpha} } \int_0 ^\infty e^{ - \left[ U(x') - U(0) \right] / k_b T } {\rm d} x' ,
\label{eqWEB07}
\end{equation}
\begin{equation}
{\partial \left( \tau_{x_0} ^{-} \right)^{\alpha} \over \partial x_{0}} |_{x_0 = 0^{-}}= -
{ 1 \over D_{\alpha} } \int_{-\infty} ^0 e^{ - \left[ U(x') - U(0) \right] / k_b T } {\rm d} x' .
\label{eqWEB08}
\end{equation}
Using Eqs. 
(\ref{eqWEB04},\ref{eqWEB07},\ref{eqWEB08})
we find
\begin{equation}
{\cal R} = { P_B ^{+} \over 1 - P_B ^{+} }, 
\label{eqratio}
\end{equation}
where $P_B ^{+}$ is Boltzmann's probability of finding the particle in
the domain $x>0$ Eq.
(\ref{eq14}).
Eqs. 
(\ref{eqWEB05},\ref{eqratio}) were found previously in \cite{Golan}
using a different approach. 
 One can show that the average occupation
time is 
\begin{equation}
\langle T^{+} \rangle \sim P_B ^{+} t 
\label{eqWEB09}
\end{equation}
and  fluctuation are very large if $\alpha<1$
\begin{equation}
\langle T^{+} \  ^2 \rangle - \langle T^{+} \rangle^2 
 \sim \left(1 - \alpha\right) P_B ^{+}  \left( 1 - P_B ^{+} \right) t^2.
\label{eqWEB10}
\end{equation}
We inrtoduce a measure for ergodicity breaking the EB parameter
\begin{equation}
{\mbox EB}={ \langle T^{+} \ ^2 \rangle - \langle T^{+} \rangle^2 \over \langle T^{+} \rangle^2} \sim \left(1 - \alpha \right){ 1 - P_B ^{+} \over  P_B ^{+}},
\end{equation}
which is zero in the ergodic phase $\alpha = 1$.

\subsection{Diffusion in an interval}

 We consider the case where the particle is free to diffuse
in an interval of total length $L^{+} + L^{-}$. The particle 
is initially on the origin $x=0$ and reflecting boundary conditions
are on $x=L^{+}$ and $x=-L^{-}$. Statistical properties of
$T^{+}$ the time spent in $(0,L^{+})$ are now investigated.

 The survival probability 
can be calculated using Eq. 
(\ref{eqfrac07})
\begin{equation}
W_{x_0} ^{+} (s) = { 1 - 
{ \cosh\left[ \sqrt{ {s^{\alpha} \over D_{\alpha} }}
 \left( L^{+} -x_0 \right) \right]\over \cosh\left( \sqrt{ {s^{\alpha} \over D_{\alpha} }}  L^{+} \right) }
\over s},
\end{equation}
and when $\alpha = 1$  we recover a text book result \cite{Redner}. 
Using Eq. (\ref{eq08}) we find 
$$ G_s (u) = $$
\begin{equation}
 { \left( s + u \right)^{ \alpha / 2 - 1} \tanh\left[ { \left( s + u \right)^{ \alpha/2} L^{+} \over \sqrt{D_{\alpha}} } \right] +
s^{ \alpha / 2 - 1} \tanh\left( { s^{ \alpha/2} L^{-} \over \sqrt{D_{\alpha}} } \right) \over 
 \left( s + u \right)^{ \alpha / 2 } \tanh\left[ { \left( s + u \right)^{ \alpha/2} L^{+} \over \sqrt{D_{\alpha}} } \right] +
s^{ \alpha / 2 } \tanh\left( { s^{ \alpha/2} L^{-} \over \sqrt{D_{\alpha}} } \right) }.
\label{eqGSU1}
\end{equation}

 For free boundary conditions, namely in the limit where
the system size is  infinite $L^{+} \to \infty$ and
$L^{-} \to \infty$ we find
\begin{equation}
G_s (u) = { \left( s + u \right)^{ \alpha/2 - 1} + s^{\alpha/2 - 1} \over
\left( s + u \right)^{\alpha/2} + s^{\alpha/2} } .
\end{equation}
Inverting to the time domain, the PDF of $T^{+}$ is the symmetric
Lamperti PDF with index $\alpha/2$
\begin{equation}
G_t   \left( T^{+} \right)={1 \over t}\delta_{\alpha/2} \left(1,{T^{+} \over t}\right).
\end{equation}
%
When $\alpha=1$, i.e. the case of normal Gaussian diffusion,
 we recover the well known arcsine distribution. As  
$\alpha$ is decreased we are more likely to find
the particle localized in
$x>0$ or $x<0$ for a time of the order of the observation time. 
Indeed when $\alpha \to 0$ the PDF of $T^{+}$ is a combination
of two delta functions with $T^{+} = t$ (particle always on $x>0$)
or $T^{+}=0$ (particle always in $x<0$).  

 A different behavior is found for finite $L^{+}$ and $L^{-}$, then
using Eq.  
(\ref{eqGSU1})
we find 
\begin{equation}
G_t \left(T^{+}  \right)  \sim 
{1 \over t} \delta_{\alpha/2} \left(1,{T^{+} \over t}  \right)  \  \ \
t<< \left[ { \mbox{min}(L^{+},L^{-})^2 \over D_{\alpha} } \right]^{ 1 /\alpha}. 
\end{equation}
For these time scales the particle does not interact with the boundaries,
and $G_t(T^{+})$ is the symmetric Lamperti PDF with index $\alpha/2$.
In the long time limit, corresponding to small $s$ $u$ limit we find
using Eq.  
(\ref{eqGSU1}) 
\begin{equation}
G_s (u) \sim { \left( s + u \right)^{\alpha - 1} L^{+} + s^{\alpha-1} L^{-} \over \left( s + u \right)^{\alpha} L^{+} + s^{\alpha} L^{-} } 
\end{equation}
and hence when $t \to \infty$
\begin{equation}
G_t \left(T^{+}  \right)  \sim 
{1 \over t} \delta_{\alpha} \left({L^{+} \over L^{-} },  {T^{+} \over t}  \right).  
\end{equation}
This is in agreement with our more general result Eqs. 
(\ref{eqWEB05},\ref{eqratio}) namely for the case of free diffusion
$P_B ^+ = L^{+}/ ( L^{+} + L^{-})$ and hence ${\cal R} = L^{+} / L^{-}$. 
If $L^{+} \ne L^{-}$ the PDF of $T^{+}$ is as expected non-symmetric, 
reflecting the tendency of the particle to reside in the larger
interval 
[say $(0,L^{+})$ if $L^{+} > L^{-}$] 
for longer times compared with the shorter domain. 
For long times an equilibrium is obtained: for $\alpha = 1$ 
an ergodic phase is found  where $T^{+} /t  = L^{+}/(L^{-} + L^{+})$
while for $\alpha<1$ weak ergodicity breaking is found. 
We see that the statistics of occupation times exhibits a transition from
a symmetric Lamperti PDF with index $\alpha/2$ when diffusion is
dominating the dynamics, i.e. for short times, to
a generally non-symmetric Lamperti PDF with index $\alpha$,
for long times when the particle interacts with the boundaries.
Such a transition is not limited to free diffusion as we show later. 

 If $L^{-} \to \infty$ while $L^{+}$ remains finite a different
behavior is found. Now the particle can be found either in
a domain of finite length $0<x<L^{+}$ or in the infinite domain 
$-\infty < x<0$. Statistically we expect of-course that the particle
will reside more in $x<0$, though the random walk is recurrent
hence, after each sojourn time in $x<0$ the particle is ejected back to
$x>0$, provided that we wait long enough. However the average 
return time from a point in $x<0$ to some point in $x>0$ is infinite,
and this means that simple scaling $T^{+} \sim t$ does not not hold. 
For this case we have
\begin{equation}
G_s(u) = { \left( s + u\right)^{\alpha/2 - 1} \tanh\left[ { \left( s + u \right)^{\alpha/2} L^{+} \over \sqrt{D_{\alpha} } } \right] + s^{\alpha/2 - 1 } \over 
\left( s + u \right)^{\alpha/2} \tanh \left[ { \left(s + u \right)^{\alpha/2} L^{+} \over \sqrt{D_{\alpha} } } \right] + s^{\alpha/2} }.
\label{eqFREE1}
\end{equation}
To investigate deviations from simple scaling we consider moments
of the random variable $T^{+}$, using the small $u$ expansion
of Eq. (\ref{eqFREE1}). The average occupation time in
$0<x<L^{+}$ is
\begin{equation}
\langle T^{+} \rangle = {\cal L}_{s \to t} ^{-1} \left[
{1 \over 2 s^2} \left( 1 - e^{ - { 2 s^{\alpha/2} L^{+} \over \sqrt{D_{\alpha} } } } \right) \right],
\end{equation}
where ${\cal L}_{s \to t} ^{-1}$ is the inverse Laplace transform.
This expression is inverted using one sided L\'evy
stable functions, recall 
\begin{equation}
 l_{\alpha/2,{ 2 L^{+}/\sqrt{D_{\alpha}}},1}(t) = {\cal L}^{-1}_{s \to t}  
e^{ - 2 s^{\alpha/2} L^{+} / \sqrt{D_\alpha}}, 
\end{equation}
and see \cite{PRE1} and Ref. therein for more mathematical
details on this function.  
The one sided stable cumulative distribution is
\begin{equation}
L_{\alpha/2,{2 L^{+}\over \sqrt{D_\alpha} }, 1}\left( t \right) = \int_0 ^t  l_{\alpha/2,{2 L^{+} \over \sqrt{D_\alpha} } , 1} \left( t \right) {\rm d} t,
\end{equation} 
and hence
\begin{equation}
\langle T^{+} \rangle = {1 \over 2} \int_0 ^t \left[ 1 - L_{\alpha/2,{2 L^{+} \over \sqrt{D_\alpha} }, 1 } \left( t \right) \right] {\rm d} t.
\end{equation}
For short times $\langle T^{+} \rangle = t/2$, since then the particle does not
have time to interact with the boundary, and it spends half of the time
in $x>0$. For long times
\begin{equation}
\langle T^{+} \rangle \sim { L^{+} \over \sqrt{D_\alpha} } {t^{1 - \alpha/2} \over \Gamma\left( 2 - \alpha/2 \right) } .
\end{equation}
We see that as the process becomes slower, namely when $\alpha$ is decreased,
the particle tends to stay more in $0<x<L^{+}$ i.e. 
$\langle T^{+} \rangle \propto t$ for 
$\alpha=0$ but $\langle T^{+}\rangle \propto t^{1/2}$ if
$\alpha=1$. We explain this result for normal diffusion
 $\alpha=1$ by thinking about the process as a two state process,
i.e. the particle is either in $x<0$ or in $x>0$. Sojourn times
in $x>0$ are finite, since the interval $0<x<L^{+}$ is finite.
The PDF of times in state $x<0$ follow the $t^{-3/2}$ power law tail due to
usual diffusion. The number of time the particle will  cross zero is  
$ \langle n(t) \rangle \sim t^{1/2}$ at-least for a lattice CTRW process
(in the continuum limit this question is not well defined). Hence
we expect $\langle T^{+} \rangle = \langle n(t) \rangle * \mbox{average time
in \ } 0<x<L^{+} \sim t^{1/2}$ as we find.  If $\alpha<1$ we expect
$\langle n(t) \rangle \sim t^{\alpha/2}$ and the average time is
$0<x<L^{+}$ is proportional to $\int^t \psi(t) t {\rm d} t\simeq 
\int^t t^{ - ( 1 + \alpha)} t {\rm d} t \simeq t^{1- \alpha}$,
hence we get $ \langle T^{+} \rangle \propto t^{1 - \alpha/2}$. 
 The second moment of $T^{+}$ is 
\begin{equation}
\langle T^{+}\ ^2 \rangle \sim 4 \left( 1 - \alpha \right) {L^{+} \over \sqrt{D_\alpha} } {t^{ 2 - \alpha/2} \over \Gamma\left( 3 - \alpha/2\right)}.
\end{equation}
Note that we do not have simple scaling and $\langle T^{+}\ ^2 \rangle \propto t^{2 - \alpha/2}$ is
not proportional to $\langle T^{+} \rangle^2 \propto t^{2 - \alpha}$,
and hence the PDF
of $T^{+}$ does not have a simple scaling. 

\subsection{Diffusion with Drift}

 We consider anomalous diffusion in the presence of 
a constant driving force $F>0$, for an infinite
system. The biased diffusion yields a net drift
$ \langle x \rangle = D_{\alpha} F t^{\alpha}/ [k_b T \Gamma( 1 + \alpha)]$.
Since $F>0$ the particle will escape to infinity, hence
for a particle starting on $x=0$ we expect $T^{+} \sim t$ when
$t$ is large. 

The survival probability in the right half space
\begin{equation}
W^{+}_{x_0} \left( s \right) = { 1 -  
\exp\left[- {F \beta^{+}(s) x_0 \over 2 k_b T}\right] \over s},
\label{eqDWD01}
\end{equation}
when $x_0>0$.
To obtain the survival probability for left random walks
replace $\beta^{+}(s)$ in Eq. (\ref{eqDWD01}) with $\beta^{-}(s)$, and 
\begin{equation}
 \beta^{\pm}(s) = 1 \pm  \sqrt{ 1 + 4 s^{\alpha} \tau^\alpha} 
\label{eqDWD02}
\end{equation}
where 
\begin{equation}
\tau^\alpha = {(k_b T)^2 \over  F^2 D_{\alpha} }.  
\end{equation}
Using Eq. 
(\ref{eq09})
\begin{equation}
G_s(u) = { {\beta^{+} (s + u) \over (s + u) } - {\beta^{-} (s) \over s} \over
\beta^{+} (s + u) - \beta^{-} (s) }.
\label{eqDWD03}
\end{equation}

For $s^{\alpha} \tau^{\alpha}>>1$ and 
$u^{\alpha} \tau^{\alpha}>>1$
\begin{equation}
G_s(u)  \simeq { (s + u)^{\alpha/2 -1 } + s^{\alpha/2 -1} \over
(s + u)^{\alpha/2} + s^{\alpha/2} }.  
\label{eqDWD04}
\end{equation}
 Thus for short times $t << \tau$ 
\begin{equation}
G_t ( T^{+} ) \simeq { 1 \over t} \delta_{\alpha/2} \left( 1 , {T^{+} \over t} \right)
\label{eqDWD05}
\end{equation}
namely a symmetric Lamperti PDF with index $\alpha/2$
describes the residence times.
Such behavior is independent of the drift and can be understood
if we notice that for short times the dynamics is governed by
diffusion not drift.  To see this recall that
the scaling of these two processes
is $x\sim t^{\alpha/2} $ (diffusion) and $x \sim t^{\alpha}$ (drift)
 and hence for
short times diffusion wins.  For long times we use the
small $s,u$ expansion of Eq. 
(\ref{eqDWD03}), $G_s(u) \sim 1/(s + u)$ which gives the
expected behavior $G_t (T^{+} ) \sim \delta ( T^{+} - t)$.

 The mean occupation time is
\begin{equation}
\langle T^{+} \rangle = {\cal L}^{-1}_{s \to t} 
\left\{ {1 \over 2 s^2} { 1 + \sqrt{ 1 + 4 s^{\alpha} \tau^{\alpha} } \over
\sqrt{ 1 + 4 s^{\alpha} \tau^{\alpha} } } \right\}.
\label{eqDWD06}
\end{equation}
For the case of normal diffusion  $\alpha=1$
we find
\begin{equation}
\langle T^{+} \rangle = {t \over 2} + \tau\left[ 
\sqrt{ { t \over \pi \tau}} e^{ - { t \over 4 \tau}} + \left( {t \over 2 \tau} - 1\right) \mbox{Erf}\sqrt{ {t \over 4 \tau} } \right].
\end{equation}
The long time behavior is
\begin{equation}
\langle T^{+} \rangle \sim t - \tau\left( 1 -   \sqrt{ { 4 \tau\over \pi t}}e^{- {t \over 4 \tau}} \right),
\end{equation}
the leading term $\langle T^{+} \rangle \sim t$ is expected since as
mentioned  for
long times the particle is always in $x>0$ when $F>0$.
For short times
\begin{equation}
\langle T^{+} \rangle \sim  { t \over 2} \left[ 1 + {2 \over 3 \sqrt{\pi}} \left({ t \over \tau}\right)^{1/2} - {1 \over 30 \sqrt{\pi}} \left( {t \over \tau}\right)^{3/2} + 0(t^{5/2})\right].
\end{equation}
The leading term $\langle T^{+}\rangle \sim  { t \over 2}$ shows that
at short time
diffusion not drift is dominating the process, hence from symmetry
half of the
time the particle is on $x>0$. 
For the sub-diffusive case $\alpha<1$ we investigate
the  long time behavior of $\langle T^{+} \rangle$ using 
the small $s$ expansion
of  
(\ref{eqDWD06}) and then inverting to the time domain
\begin{equation}
\langle T^{+} \rangle \sim t\left[1 - {1 \over \Gamma\left(2- \alpha\right)} {(k_b T)^2 \over F^2 D_{\alpha} t^\alpha}  +
\mbox{O}\left({\tau \over t}\right)^{2 \alpha} \right].
\end{equation}
At short times we use Eq. (\ref{eqDWD06}) and
 Hankel's contour integral, for the $\Gamma(z)$
function, and find
\begin{equation}
\langle T^+ \rangle \sim {t \over 2}  \left[ 1 + { 1 \over 2 \Gamma\left( 2 + \alpha/2\right)} { F \sqrt{D_{\alpha}}  \over \left( k_b T \right)^2} t^{\alpha/2}  + \cdots\right].
\end{equation}

 We note that results for the case $F<0$ can be easily 
obtained from our results for $F>0$. The distribution of
times $T^{-}$ in $x<0$ when $F>0$ is equal of-course to the
distribution of time $T^{+}$ in $x>0$ when $F<0$. Also
$T^{+} + T^{-} = t$ hence a simple shift of the random variable
yields $T^{-}=t-T^{+}$, and hence also statistics for the case $F<0$.

\subsection{Diffusion in an Unstable Force Field}

 We consider a particle in an unstable force field
\begin{equation}
F(x) = \left\{ 
\begin{array}{c c}
F_{+} & \  x>0  \\
- F_{-} & \  x<0  
\end{array}
\right.
\end{equation}
where $F_{+} > 0$ and $F_{-} > 0$.
For this case the particle will eventually escape either to
$+\infty$ or $-\infty$, and the random walk is not recurrent. 
The survival probabilities in left and right domains are
\begin{equation}
W_{x_0} ^{\pm} (s ) = {1 \over s} \left\{  1 - \exp\left[ \mp F_{\pm} \left( 1 + \sqrt{ 1 + 4 s^{\alpha} \tau_{\pm} ^{\alpha} } \right) x_0/ \left(2 k_b T\right) \right] \right\}.
\label{eqUS01}
\end{equation}
where
\begin{equation}
\tau_{\pm} ^{\alpha} = { (k_b T)^2 \over F_{\pm} ^2 D_\alpha}.
\end{equation}
Using Eq. (\ref{eq08})  we find
\begin{equation}
G_s(u)  = { { \gamma_{+}  \left(s + u \right) \over s + u } + {\gamma_{-}(s) \over s} \over \gamma_{+} \left(s + u\right) + \gamma_{-}\left(s\right) }
\label{eqUS02}
\end{equation}
where
\begin{equation}
\gamma_{\pm} (s) = F_{\pm}\left( 1 + \sqrt{1 + 4 s^\alpha \tau_{\pm} ^{\alpha} } \right). 
\end{equation}

 Using the definition Eq.
(\ref{eq11a}),
the small $s$ behavior of  
Eq. (\ref{eqUS01}) gives the survival probabilities in the $+$ and $-$
domains, $x>0$ and $x<0$ respectively.
In the limit of long times 
\begin{equation} 
\epsilon_{\pm} (x_0)= 1 - \exp\left( \mp {F_{\pm} x_0 \over k_b T} \right),
\end{equation}
where $x_0>0$ for $+$ and $x_0<0$ for $-$. 
Hence according to the rather general Eqs. 
(\ref{eq14a},\ref{eq14b}) 
we find
\begin{equation}
G_t(T^{+} ) \sim { F_{+} \over F_{-} + F_{+} } \delta\left( T^{+} - t \right) + {F_{-} \over F_{+} + F_{-} } \delta(T^+ ) .
\label{eqUS03}
\end{equation}
This long time behavior exhibits the same behavior for the normal
diffusion $\alpha=1$ as for the anomalous case $\alpha<1$. 
Note that Eqs. (\ref{eq14a},\ref{eq14b}) where derived for normal
diffusion however one can show that they are valid also for the
anomalous diffusion case. To see this use the small $s,u$ expansion of
Eq.  
(\ref{eqUS02}) which gives
\begin{equation}
G_s(u) \sim { 1 \over F_{+} + F_{-} }  \left( {F_{+} \over s+u  } + {F_{-} \over s} \right),
\end{equation}
which is the double inverse Laplace transform of Eq. 
(\ref{eqUS03}) and is independent of the parameters 
$\alpha,D_{\alpha}$ and $T$. 

For short times $t\ll \mbox{min}\left( \tau_{-},\tau_{+} \right)$
we use the large $s,u$ behavior of Eq.
(\ref{eqUS02}) and find $G_t(T^{+}) \simeq {1 \over t} \delta_{\alpha/2} \left( 1 , {T^{+} \over t}\right)$. Thus for short times the PDF
of $T^{+}$ is the symmetric Lamperti PDF which is independent
of all the parameters of the problem except for $\alpha$. At the early stages
of the dynamics the  
diffusion process not the drift is the most important,
and  hence
forces are not relevant. 

\subsection{Diffusion in Binding Force Field}

 We consider a particle in a stable force field
\begin{equation}
F(x) = \left\{ 
\begin{array}{c c}
-F_{+} & \  x>0  \\
 F_{-} & \  x<0  
\end{array}
\right.
\end{equation}
where $F_{+} > 0$ and $F_{-}>0$. The random walk is recurrent. Using
an approach similar to one used in previous subsection
\begin{equation}
G_s(u) = { {\xi_{+}\left( s + u \right) \over s + u } + {\xi_{-} \left( s \right) \over s} \over \xi_{+}\left( s + u \right) + \xi_{-} \left( s \right) } ,
\end{equation}
and
$\xi_{\pm}\left(s\right)=F_{\pm}\left(1-\sqrt{1+4s^\alpha \tau_{\pm} ^\alpha}\right)$. 

 For small $s,u$ we have
\begin{equation}
G_s (u) \sim { \left( s + u \right)^{\alpha - 1} {F_{-} \over F_{+}} + s^{\alpha - 1} \over \left( s + u \right)^{\alpha} { F_{-} \over F_{+} } + s^{\alpha} },
\end{equation} 
and hence when $t$ is large 
\begin{equation}
G_t\left( t^{+} \right) \sim {1 \over t} \delta_\alpha \left( {F_- \over F_+}, {T^{+} \over t } \right). 
\end{equation}
This is in agreement with our more general results Eqs.
(\ref{eqWEB05},
\ref{eqratio}). For short times we have 
$G_t(T^{+}) \simeq {1 \over t} \delta_{\alpha/2} \left( 1 , {T^{+} \over t}\right)$ which is similar to the behavior of the unstable
field discussed in previous sub-section.

\section{Discussion}

 Statistics of occupation times for {\em  binding external fields}
exhibits in the limit
of long times an ergodic behavior when the diffusion is normal, or
weak ergodicity breaking Eqs. 
(\ref{eqWEB05},\ref{eqratio})
when diffusion is anomalous. We established
a link between weak ergodicity breaking and fractional calculus.
The exponent $\alpha$ in the
fractional derivative  $\partial^\alpha / \partial  t^\alpha$
enters in Eq. (\ref{eqWEB05}) describing the non-ergodic properties of 
the residence times.  Since many processes and systems are modeled today using 
the fractional calculus approach, it is not out of the question that
weak ergodicity breaking has many applications, and is wide spread.
We can say that at-least 
one must treat with care, results obtained using fractional
kinetic equations, since they describe only ensemble averages, not time
averages. 

 For binding external fields our results are in full agreement
with those derived recently, by Bel and the author \cite{Golan}. 
There a continuous
time random walk process was considered. Technically the methods
used to treat the two problems are different. For the fractional 
framework, a differential equation, Eq. 
(\ref{eqfrac02})
 for the occupation times
is derived and  solved, which yields the weakly non ergodic properties of
the system, while for the CTRW certain recursive relations
must be solved \cite{Golan1}. Our work shows how the fractional framework, which
is the continuum limit of the CTRW (and in this sense simpler)
can be used to obtain statistics of residence times,
and for binding fields weak ergodicity breaking. 

 We found a general relation between the problem of occupation times
and the problem of first passage times, Eq. (\ref{eq08}). 
Mathematically the problem of first passage time is 
described in terms of a differential equation whose solution depends
on a single time $t$ as a parameter, while the equation
for residence time  depends on
two times $T^{+}$ and $t$, hence the approach based on 
calculation of first passage times  instead of a direct
calculation of distribution of residence times seems to us useful.
Besides, statistics of first passage times in external
fields is  a well investigated problem and this knowledge
was used here.  
We showed that from Eq. (\ref{eq08}) general properties
of residence times can be easily derived. 
 Luckily the  relation Eq.
(\ref{eq08}) is valid for both the non-Markovian sub-diffusive and normal
processes, and hence has some general validity. From our
derivation of this relation, we see that a key ingredient for
this relation to be valid is the renewal property of the underlying
random walk. As far as we know
statistics of  residence times  for
more general non-Markovian and non renewal dynamics  
was not investigated so far \cite{Dhar2}.

 As mentioned, for binding fields we confirm old results
in the limit of long measurement times. 
 New results for
non-binding force fields and also for non
asymptotic times are also found. 
For example, for the
simplest case of  anomalous
diffusion in free space, we found that the PDF of
occupation times in half space is a symmetric Lamperti
PDF with index $\alpha/2$. This is the natural generalization
of the well known arcsine law. Hence  we showed that the fractional
kinetic framework is indeed a natural generalization of the 
ordinary diffusion
process. 

 Other  behaviors are found when diffusion is
not free. For example for short times
we expect rather generally that occupation times statistics
is described by a symmetric Lamperti PDF, since for short times
drift by force fields is a slow process if compared with the diffusion
process.
For long times we get either (i) weak ergodicity breaking if
the force field is binding, or (ii) occupation times described
by Eq. 
(\ref{eq14a})
if both the left and right random walks are not recurrent
(note that this equation is also valid for anomalous sub diffusion),
or (iii)  when random walk in the left and right domain is not bounded and
recurrent, in such a way that the average first return time
is infinite,  then a Lamperti PDF describes statistics 
of occupation
times in the limit of long times. For normal
diffusion process the asymmetry parameter is determined
by Eq. 
(\ref{eqAsym}).
We also analyzed  other cases where 
the left random walk is recurrent but not bounded
and the right random walk is bounded, showing
deviations from simple scaling.

{\bf Acknowledgment} 
The  Israel Science Foundation supported this work.
I thank A. Comtet, S. Majumdar and  G. Margolin,  
for comments on the manuscript, and
Golan Bel for fruitful  discussions and
a previous collaboration, which led to the current research.

\section{Appendix A} 

 The Fokker--Planck equation for a particle in a force
field reads
\begin{equation}
{\partial c(x,t) \over \partial t} = D_1 L_{{\rm FP}} c(x,t)
\label{eqA1}
\end{equation}
with the operator
\begin{equation}
L_{{\rm FP}} = {\partial^2 \over \partial x^2} - {\partial \over \partial x} {F(x) \over k_B  T},
\end{equation}
where $D_1$ is the diffusion coefficient. 
To prepare for the FFPE we rewrite Eq. (\ref{eqA1}) in
an integral form
\begin{equation}
c(x,t) - \delta(x-x_0) = D_1 \ _0 I_t ^1 L_{{\rm FP}} c(x,t)
\end{equation}
where $\delta(x-x_0)$ are the initial conditions. According
to the fractional kinetic approach, we must replace the integral
$\ _0 I_t ^1$ with a fractional Riemann-Liouville integration, defined
as an operation on a function $Z(t)$ according to
\begin{equation}
\ _0 I_t ^{\alpha} Z(t) \equiv { 1 \over \Gamma(\alpha)} \int_0 ^t { Z(t') \over \left( t - t'\right)^{1 - \alpha} } {\rm d} t',
\end{equation}
and for our purpose  $0<\alpha<1$. The FFPE in its integral form is
\begin{equation}
c(x,t) - \delta(x-x_0) = D_\alpha \ _0 I_t ^\alpha L_{{\rm FP}} c(x,t)
\label{eqA2}
\end{equation}
where $D_{\alpha}$ is a generalized diffusion coefficient. 
Several authors present this equation in different ways.
Sometimes \cite{MBK}  further differentiation with respect to
time is made  in Eq. (\ref{eqA2}) to return to a fractional
 differential equation
instead of the fractional integral form.
However, following work of Gorenflo and
 Mainardi  e.g. \cite{Gor} there is now growing use of Caputo 
symbols which are more elegant. Such symbols are used in
Eq. (\ref{eqFFPE})
which has the same meaning as  
Eq. (\ref{eqA2}).  
Later we use the Laplace $t\to s$ 
 transform of the FFPE equation
(\ref{eqA2})
\begin{equation}
s c(x,s) - \delta(x- x_0) = D_{\alpha} s^{1 - \alpha} L_{{\rm FP}} c(x,s).
\label{eqA3}
\end{equation}
We see that the solution of FFPE with $\alpha<1$ is related to the
solution of the Fokker-Planck equation when $\alpha=1$ in Laplace
space. To obtain the solution of the FFPE from the solution
of the usual Fokker-Planck equation we must make a replacement
$D_1 \to D_{\alpha} s^{1 - \alpha}$ \cite{MBK}. 
This similarity transformation
in $s$ space can be inverted to real time \cite{PRE1}, and  with it
one can obtain a solution of the FFPE once the corresponding
solution of the usual Fokker-Planck equation is known.
This transformation is related to subordination and the inverse L\'evy
transform \cite{PRE1}.

 Similarly  the survival probability  for the normal diffusion
case, and the fractional case are related to each other, by 
a simple transformation  in Laplace space \cite{PRE1}.
According to Eq. (57) in Ref. \cite{PRE1}
\begin{equation}
 W^+_{\alpha,x_0}(s)= {D_1\over D_{\alpha}} s^{\alpha -1}W^+_{1,x_0}\left({D_1 \over D_\alpha} s^{\alpha} \right),
\label{eqAp01}
\end{equation}
where $W^+_{\alpha,x_0}(s)$ is the survival probability for
the fractional $\alpha<1$ or the normal case $\alpha=1$.
Using Eqs. 
(\ref{eq07},
\ref{eqAp01})
it is easy to prove the validity of Eq. (\ref{eqfrac07}).

\end{document}